\documentclass[12pt]{article}
\usepackage{graphicx}
\newcommand{\dbox}{\,\raise2pt\hbox{\fbox{\rule{2.5pt}{0pt}\rule{0pt}{2.5pt}}}\,}
\newcommand{\qed}{\,\raise0pt\hbox{\mbox{\rule{6.5pt}{6.5pt}}}}

\begin{document}
\setlength{\baselineskip}{7mm}


\begin{titlepage}
 \begin{normalsize}
  \begin{flushright}
        UT-Komaba/18-4\\ KOBE-TH-18-03\\ EHIME-TH-102\\
        September 2018
  \end{flushright}
 \end{normalsize}
 \begin{LARGE}
   \vspace{1cm}
   \begin{center}
     A lattice formulation of \\${\cal N}=2$ supersymmetric SYK model
   \end{center}
 \end{LARGE}
 \vspace{5mm}
 \begin{center}

    Mitsuhiro {\sc Kato}$^{a}$, 
            \hspace{3mm}
    Makoto {\sc Sakamoto}$^{b}$ 
            \hspace{3mm}and\hspace{3mm}
    Hiroto {\sc So}$^{c}$ 
    \\
      \vspace{4mm}
        ${}^{a}${\sl Institute of Physics} \\
        {\sl University of Tokyo, Komaba}\\
        {\sl Meguro-ku, Tokyo 153-8902, Japan}\\
       \vspace{4mm}
        ${}^{b}${\sl Department of Physics}\\
        {\sl Kobe University}\\
        {\sl Nada-ku, Hyogo 657-8501, Japan}\\
       \vspace{4mm}
        ${}^{c}${\sl Department of Physics}\\
        {\sl Ehime University}\\
        {\sl Bunkyou-chou 2-5, Matsuyama 790-8577, Japan}\\
     \vspace{1cm}

  ABSTRACT\par
 \end{center}
 \begin{quote}
  \begin{normalsize}

We construct ${\cal N}=2$ supersymmetric SYK model on one-dimensional (euclidean time) lattice.
One nilpotent supersymmetry is exactly realized on the lattice in use of the cyclic Leibniz rule (CLR).

  \end{normalsize}
 \end{quote}

\end{titlepage}
\vfil\eject


Sachdev-Ye-Kitaev (SYK) model\cite{Sachdev:1992fk,Kitaev} and its generalizations have been attracting much attentions in several contexts.\footnote{For  recent review see, for example, \cite{Rosenhaus:2018dtp, Sarosi:2017ykf} and references therein.} In these analysis, large $N$ limit is fully utilized to obtain effective theory in the IR region. Finite $N$ analysis, however, is also interesting to see, for example, some sort of stringy corrections for AdS/CFT-type correspondence.

For finite $N$ analysis beyond perturbation, one approach is the exact diagonalization of the hamiltonian\cite{Li:2017hdt} based on the realization of fermion operators by gamma matrices. Another approach may be the lattice formulation on the euclidean time. The latter seems convenient for the calculation of multi-point correlation function of the operators with distinct times, especially for the comparison with results from the effective field theory of bi-local collective modes. Also, if higher-dimensional extension becomes possible, Monte Carlo simulation will be numerically less expensive than exact diagonalization of the hamiltonian. So we concentrate lattice formulation in the present paper.
Among others we will focus on lattice formulation of the supersymmetric generalization\cite{Fu:2016vas} of SYK model. It is actually highly non-trivial, since realizing supersymmetry on lattice is a very difficult task\cite{Kato:2008sp}.

The present authors have been studying a way for realizing nilpotent subalgebra of the supersymmetry on lattice in use of the cyclic Leibniz rule (CLR)\cite{Kato:2013sba, Kato:2016fpg}. In the present letter, as an application of the CLR, we will construct ${\cal N}=2$ supersymmetric SYK model\cite{Fu:2016vas} on lattice. As will be seen, one of the two supersymmetries is exactly realized on lattice thanks to the CLR.

Let us consider ${\cal N}=2$ supersymmetric SYK model\cite{Fu:2016vas} whose hamiltonian $H$ is given by the anti-commutator of two nilpotent supercharges $Q$ and $\bar Q$
\begin{equation}
H=\{Q\,,\bar Q\},\qquad Q^2=0,\qquad \bar Q^2=0,
\end{equation}
where supercharges are defined by $N$ complex fermions $\psi^i, \bar\psi^i$ $(i=1,\cdots,N)$ and complex random couplings $C_{ijk}, \bar C_{ijk}$ with totally anti-symmetric indices:
\begin{equation}
Q=\frac{i}{3!}C_{ijk}\psi^i\psi^j\psi^k,\qquad
\bar Q=\frac{i}{3!}\bar C_{ijk}\bar\psi^i\bar\psi^j\bar\psi^k.
\end{equation}
These fermions satisfy the following anti-commutation relations
\begin{equation}
\{\psi^i\,,\psi^j\}=0,\qquad
\{\bar\psi^i\,,\bar\psi^j\}=0,\qquad
\{\psi^i\,,\bar\psi^j\}=\delta^{ij}.
\end{equation}
The random couplings have non-zero second moment under quenched average
\begin{equation}
\overline{C_{ijk}\bar C_{ijk}}=\frac{2J}{N^2}
\end{equation}
with a characteristic constant $J$ which controls the strength of the interaction.

The above supercharges are cubic in fermion, thereby hamiltonian has quartic interactions. This can be generalized by taking any odd number of fermions in supercharges. Hereafter we use generalized supercharges with $q$ fermions.
Corresponding action in euclidean time is defined by
\begin{equation}
S=\int dt \left\{ \bar{\psi}^i\partial_t\psi^i-\bar{b}^ib^i
+\frac{i^{\frac{q-1}{2}}}{(q-1)!}C_{j_1j_2\cdots j_q}\bar{b}^{j_1}\psi^{j_2}\cdots\psi^{j_q}
+\frac{i^{\frac{q-1}{2}}}{(q-1)!}\bar C_{j_1j_2\cdots j_q}b^{j_1}\bar\psi^{j_2}\cdots\bar\psi^{j_q}
 \right\}.\label{action1}
\end{equation}
Here we have introduced complex auxiliary variables $b^i$, $\bar b^i$ in order to realize supersymmetry linearly and make the action off-shell invariant. $C$ and $\bar C$ are $q$ indices generalization of the random couplings.
Supersymmetry transformation for each variable is
\begin{equation}
\begin{array}{ccl}
\delta_Q\psi^i&=&0,\\[5pt]
\delta_Q b^i&=&0,\\[5pt]
\delta_Q\bar\psi^i&=& b^i,\\[5pt]
\delta_Q\bar b^i&=&\partial_t\psi^i,
\end{array}
\qquad\qquad\quad
\begin{array}{ccl}
\delta_{\bar Q}\psi^i &=&\bar b^i,\\[5pt]
\delta_{\bar Q} b^i&=&\partial_t\bar\psi^i,\\[5pt]
\delta_{\bar Q}\bar\psi^i&=&0,\\[5pt]
\delta_{\bar Q}\bar b^i&=&0,
\end{array}
\label{trans1}
\end{equation}
where we omit transformation parameters, so that $\delta_Q$ and $\delta_{\bar Q}$ should be treated as Grassmann odd quantities.

Now let us make a lattice version of (\ref{action1}) and (\ref{trans1}).
First, we replace the variables $\psi^i(t)$, $\bar\psi^i(t)$, $b^i(t)$ and $\bar b^i(t)$ by the lattice variables $\psi^i_n$, $\bar\psi^i_n$, $b^i_n$ and $\bar b^i_n$ where $n$ stands for a lattice site. Then supersymmetry transformation should become
\begin{equation}
\begin{array}{ccl}
\delta_Q\psi^i_n&=&0,\\[5pt]
\delta_Q b^i_n&=&0,\\[5pt]
\delta_Q\bar\psi^i_n&=& b^i_n,\\[5pt]
\delta_Q\bar b^i_n&=&(\nabla^{\rm (T)}\psi^i)_n,
\end{array}
\qquad\qquad
\begin{array}{ccl}
\delta_{\bar Q}\psi^i_n &=&\bar b^i_n,\\[5pt]
\delta_{\bar Q} b^i_n&=&(\nabla^{\rm (T)}\bar\psi^i)_n,\\[5pt]
\delta_{\bar Q}\bar\psi^i_n&=&0,\\[5pt]
\delta_{\bar Q}\bar b^i_n&=&0,
\end{array}
\label{trans2}
\end{equation}
where $\nabla^{\rm (T)}$ is an appropriate difference operator. We use superscript ${}^{\rm (T)}$ in order to distinguish it from the difference operator in the action for which we use $\nabla^{\rm (A)}$.

Next, we construct a lattice action in the following form
\begin{eqnarray}
S\,\, &=&S_{\rm kin} + S_{M} +S_{\bar M},
\label{action2}\\
S_{\rm kin}\!\!&=&\bar\psi^i_n\nabla^{\rm (A)}_{nm}\psi^i_m-\bar b^i_nb^i_n,\\
S_{M}&=&\frac{i^{\frac{q-1}{2}}}{(q-1)!}C_{j_1j_2\cdots j_q}M_{n_1n_2\cdots n_q}\bar{b}^{j_1}_{n_1}\psi^{j_2}_{n_2}\cdots\psi^{j_q}_{n_q},\\
S_{\bar M}&=&\frac{i^{\frac{q-1}{2}}}{(q-1)!}\bar C_{j_1j_2\cdots j_q}\bar M_{n_1n_2\cdots n_q}b^{j_1}_{n_1}\bar\psi^{j_2}_{n_2}\cdots\bar\psi^{j_q}_{n_q},
\end{eqnarray}
where repeated lattice site indices are summed. $M_{n_1n_2\cdots n_q}$ and $\bar M_{n_1n_2\cdots n_q}$ are complex coefficients which define multiple product of variables. Note that the last $q-1$ site indices of $M$ and $\bar M$ are  totally symmetric. This action gives (\ref{action1}) in the naive continuum limit as far as $M$ and $\bar M$ go to 1 and $\nabla$ goes to $\partial_t$. Requiring the invariance of the action under either transformation $\delta_Q$ or $\delta_{\bar Q}$ in (\ref{trans2}), we obtain the conditions which should be satisfied by $\nabla^{\rm (A)}$, $\nabla^{\rm (T)}$, $M$ and $\bar M$. For example if we require $\delta_{\bar Q}$-invariance, then we have
\begin{eqnarray}
\delta_{\bar Q} S_{\rm kin}&=&-\bar\psi^i_n(\nabla^{\rm (A)}_{nm}+\nabla^{\rm (T)}_{mn})\bar b^i_m \,\,=0,\label{inv1}\\
\delta_{\bar Q}S_{M}&=&\frac{i^{\frac{q-1}{2}}}{(q-1)!}C_{j_1j_2\cdots j_q}M_{n_1n_2\cdots n_q}
\sum_{k=2}^q(-1)^{k-2}\bar b^{j_1}_{n_1}\psi^{j_2}_{n_2}\cdots\psi^{j_{k-1}}_{n_{k-1}}
\bar b^{j_k}_{n_k}\psi^{j_{k+1}}_{n_{k+1}}\cdots\psi^{j_q}_{n_q} \,\,=0,\label{inv2}\\
\delta_{\bar Q}S_{\bar M}&=&\frac{i^{\frac{q-1}{2}}}{(q-1)!}\bar C_{j_1j_2\cdots j_q}\bar M_{mn_2\cdots n_q}\nabla^{\rm (T)}_{mn_1}\bar\psi^{j_1}_{n_1}\bar\psi^{j_2}_{n_2}\cdots
\bar\psi^{j_q}_{n_q} \,\,=0.\label{inv3}
\end{eqnarray}
The first condition (\ref{inv1}) is satisfied if difference operator $\nabla$ meets
\begin{equation}
\nabla^{\rm (A)}_{nm}+\nabla^{\rm (T)}_{mn}=0.
\label{symdiff}
\end{equation}
The second condition (\ref{inv2}) is satisfied if we make
\begin{equation}
 \mbox{$M$ to be totally symmetric for all $q$ indices.} \label{symmetric}
\end{equation}
And the third condition (\ref{inv3}) is satisfied if $\bar M$ meets
\begin{equation}
\sum_{\mbox{\scriptsize permutation of }\{n_1\cdots n_q\}}\nabla^{\rm (T)}_{mn_1}\bar M_{mn_2\cdots n_q}=0.\label{CLR1}
\end{equation}
The summation in (\ref{CLR1}) can be reduced to the one in only cyclic permutation due to the totally symmetric nature of the last $q-1$ indices.
Thus this relation is nothing but the cyclic Leibniz rule (CLR)\cite{Kato:2013sba, Kato:2016fpg}.

It should be stressed that the CLR (\ref{CLR1}) is not an abstract relation but has many concrete solutions. For example, if we simply take $\nabla^{\rm (A)}=\nabla^{\rm (T)}$, then
\begin{eqnarray}
M_{lmn}&=&\delta_{l,m}\delta_{l,n},\\
\bar M_{lmn}&=&\frac{1}{6}\left(2\delta_{l,m-1}\delta_{l,n-1}+\delta_{l,m+1}\delta_{l,n-1}+\delta_{l,m-1}\delta_{l,n+1}+2\delta_{l,m+1}\delta_{l,n+1}\right),\\
\nabla_{mn}&=&\frac{1}{2}\left(\delta_{m+1,n}-\delta_{m-1,n}\right)\label{symdiff1}
\end{eqnarray}
is one of the ultralocal\footnote{Here ultralocal means the operator defined in a region with finite extent on a lattice. Local operators on the lattice include not only ultralocal ones but also ones with infinite extent which decay exponentially.} solutions of (\ref{symdiff}), (\ref{symmetric}) and (\ref{CLR1}) for $q=3$. Systematic construction of the solutions for the CLR with symmetric difference operator can be found in \cite{Kadoh:2015zza}.

Although this might be the simplest solution, we would have species doubler with it. So we propose alternative solution
\begin{eqnarray}
M_{lmn}&=&\delta_{l,m}\delta_{l,n},\\
\bar M_{lmn}&=&\frac{1}{24}\left[2(1+r)^2\delta_{l+1,m}\delta_{l+1,n}
+2(1-r)^2\delta_{l-1,m}\delta_{l-1,n}\right.\nonumber\\
&&\left.+(1-r^2)(\delta_{l-1,m}\delta_{l+1,n}+\delta_{l+1,m}\delta_{l-1,n})\right.\nonumber\\
&&\left.+(3-r)(1+r)(\delta_{l+1,m}\delta_{l,n}+\delta_{l,m}\delta_{l+1,n})\right.\nonumber\\
&&\left.+(3+r)(1-r)(\delta_{l-1,m}\delta_{l,n}+\delta_{l,m}\delta_{l-1,n})\right.\nonumber\\
&&\left.+2(3+r^2)\delta_{l,m}\delta_{l,n}\right], \\
\nabla^{\rm (T)}_{mn}&=&\frac{1}{2}\left[\delta_{m+1,n}-\delta_{m-1,n}
+r(\delta_{m+1,n}+\delta_{m-1,n}-2\delta_{m,n})\right],\\
\nabla^{\rm (A)}_{mn}&=&\frac{1}{2}\left[\delta_{m+1,n}-\delta_{m-1,n}
-r(\delta_{m+1,n}+\delta_{m-1,n}-2\delta_{m,n})\right].
\end{eqnarray}
Here $r$ is a real parameter and the kinetic action with this $\nabla^{\rm (A)}$ contains so-called Wilson term ($r$ corresponds to the Wilson term coefficient) which lifts doublers up with cutoff-scale mass.

For $r=1$ case, which corresponds to the forward difference operator, we can write down solutions with generic $q$:
\begin{eqnarray}
\quad M_{l\,n_1\cdots n_{q-1}}&&\nonumber\\
&&\hspace{-15mm}=\frac{1}{q!}\sum_P\left(
\delta_{l+1,n_1}\cdots\delta_{l+1,n_{q-1}}+
\delta_{l+1,n_1}\cdots\delta_{l+1,n_{q-2}}\delta_{l,n_{q-1}}+
\delta_{l+1,n_1}\cdots\delta_{l+1,n_{q-3}}\delta_{l,n_{q-2}}\delta_{l,n_{q-1}}
\right.\nonumber\\
&&\left.+\cdots+
\delta_{l+1,n_1}\delta_{l,n_2}\cdots\delta_{l,n_{q-1}}+
\delta_{l,n_1}\cdots\delta_{l,n_{q-1}}\right)\\
&&\hspace{-15mm}=\frac{1}{q!}\sum_P\sum_{k=0}^{q-1}
\left[\prod_{a=1}^k\delta_{l+1,n_a}\right]
\left[\prod_{b=k+1}^{q-1}\delta_{l,n_b}\right]
\end{eqnarray}
where $P$ stands for the summation over all permutations of $(n_1,n_2,\cdots,n_{q-1})$, and $\displaystyle\prod_{a=1}^{0}\delta_{l+1,n_a}=1=\prod_{b=q}^{q-1}\delta_{l,n_b}$ is understood.

Thus we have a concrete way how to construct $\delta_{\bar Q}$-invariant lattice action for ${\cal N}=2$ supersymmetric SYK model. If you need $\delta_Q$-invariant action, just exchange the roles of $M$ and $\bar M$.

A few remarks are in order:

We cannot require both $\delta_Q$ and $\delta_{\bar Q}$ invariances. Because these impose simultaneously the CLR relation and totally symmetric indices to $M$ and $\bar M$, but there is no local solution of CLR with symmetric $M$ or $\bar M$\cite{Kato:2013sba}.

For $\delta_{\bar Q}$-invariance $M$ is totally symmetric, but $\bar M$ is not because of CLR and locality. Therefore $\bar M$ is not complex conjugate of $M$, so that the resulting action is not hermitian. This ``could-be sign problem" is of order $O(a)$ where $a$ is lattice constant, and disappears at least in the naive continuum limit.

It seems the CLR approach is a unique way to realize supersymmetry for the models of this type. The other approaches, like a method through the Nicolai map or a method of finding a nilpotent transformation without difference operator\cite{Catterall:2001fr, Catterall:2003wd, Giedt:2004qs}, may not work for the model. In particular, the last term of the action (\ref{action1}) is $\delta_{\bar Q}$-invariant but not $\delta_{\bar Q}$-exact, therefor topological field theoretic approach cannot be applied. This supports the uniqueness of the CLR approach. 

\section*{Acknowledgements}
This work is supported in part by the Grant-in-Aid for Scientific 
Research No.25287049 (M.K.), No.15K05055 (M.S.) and No.25400260 (H.S.)
by the Japanese Ministry of Education, Science, Sports and Culture. 



\end{document}